# Aluminum Oxide at the Monolayer Limit *via* Oxidant-free Plasma-Assisted Atomic Layer Deposition on GaN

*Alex Henning[§, ⊥, *], Johannes D. Bartl[§, #, ⊥], Andreas Zeidler[§], Simon Qian[‡], Oliver Bienek[§], Chang-Ming Jiang[§], Claudia Paulus[§], Bernhard Rieger[#], Martin Stutzmann[§], Ian D. Sharp[§, *]*

[§]Walter Schottky Institute and Physics Department, Technical University of Munich, 85748 Garching, Germany
[#]WACKER-Chair of Macromolecular Chemistry, Catalysis Research Center, Technical University of Munich, 85748 Garching, Germany
[‡]Department of Chemistry, Technical University of Munich, 85748 Garching, Germany

[⊥] **These authors contributed equally.**
[*] E-mail: sharp@wsi.tum.de, alex.henning@wsi.tum.de

**Abstract:** Atomic layer deposition (ALD) is an essential tool in semiconductor device fabrication that allows the growth of ultrathin and conformal films to precisely form heterostructures and tune interface properties. The self-limiting nature of the chemical reactions during ALD provides excellent control over the layer thickness. However, in contrast to idealized growth models, it is experimentally challenging to create continuous monolayers by ALD because surface inhomogeneities and precursor steric interactions result in island growth during film nucleation. Thus, the ability to create pin-hole free monolayers by ALD would offer new opportunities for controlling interfacial charge and mass transport in semiconductor devices, as well as for tailoring surface chemistry. Here, we report full encapsulation of *c*-plane gallium nitride (GaN) with an ultimately thin (~3 Å) aluminum oxide ($AlO_x$) monolayer, which is enabled by the partial conversion of the GaN surface oxide into $AlO_x$ using a combination of trimethylaluminum deposition and hydrogen plasma exposure. Introduction of monolayer $AlO_x$ significantly modifies the physical and chemical properties of the surface, decreasing the work function and introducing new chemical reactivity to the GaN surface. This tunable interfacial chemistry is highlighted by the reactivity of the modified surface with phosphonic acids under standard conditions, which results in self-assembled monolayers with densities approaching the theoretical limit. More broadly, the presented monolayer $AlO_x$ deposition scheme can be extended to other dielectrics and III-V-based semiconductors, with significant relevance for applications in optoelectronics, chemical sensing, and (photo)electrocatalysis.

# 1. Introduction

Conformal and sub-nanometer thin dielectric layers can be grown by atomic layer deposition (ALD) at large scale and are essential for semiconductor device applications, including for gate dielectrics in field-effect transistors,[1] carrier-selective contacts in solar cells,[2] and corrosion protection layers in (photo)electrochemical cells,[3] as well as for sensing and catalysis.[4] While the properties of ALD films can be precisely controlled by substrate surface preparations, precursor chemistries, and external process parameters,[5] complex physical and chemical interactions lead to film and interface non-idealities. As a prominent example of this, ALD is often considered to proceed *via* layer-by-layer growth since the available surface binding sites react with a gas-phase reactant until saturation is reached, ideally resulting in the formation of one monolayer with every cycle. However, a combination of precursor steric effects, adsorption energetics and substrate surface inhomogeneities result in incomplete surface coverage during film nucleation. As a consequence, islands form during the nucleation phase and growth of additional layers on already formed islands is unavoidable during subsequent ALD cycles. Thus, in contrast to the idealized concept of layer-by-layer growth, it is extremely challenging to create pin-hole free films at the monolayer limit by ALD. Overcoming this gap by growing continuous and conformal monolayers offers significant opportunities for creating atomically abrupt heterostructures, as well as for tailoring the electronic properties and chemistries of surfaces to achieve controlled functionalization or passivation.

In this work, we report the formation of an ultimately thin (~3 Å), yet continuous, aluminum oxide ($AlO_x$) monolayer on gallium nitride (GaN) using an oxidant-free ALD process. To the best of our knowledge, this is the first experimental report of the formation of a closed single monolayer of $AlO_x$ on a non-metal surface. Although the strategy can be extended to other materials, GaN was selected due to its established technological importance. In particular, GaN is a III-V compound semiconductor that is industrially relevant for high frequency and high-power electronics because of its large breakdown field, thermal conductivity, thermal stability, and mobility.[6] In addition, it has a direct band gap (3.4 eV) and is broadly used in modern optoelectronics applications. However, a major challenge of III-V semiconductor technology is to precisely engineer interface properties, which is often hampered by high concentrations of surface states. This issue is especially pronounced for polar GaN surfaces due to polarization-compensating surface charges. In this respect, dielectric passivation by ALD is promising since it has been demonstrated to significantly reduce interface state densities within GaN devices.[7] A major advantage of ALD is its ability to synthesize defined films at low temperature, which provides versatile process compatibility and synthetic access to amorphous dielectrics that have practical advantages as passivation and interfacial layers.[8] The most widely utilized and industrially relevant ALD process involves trimethylaluminum (TMA) and an oxidant (typically water), sequentially introduced into a reactor chamber, to form amorphous $AlO_x$. TMA is an electron-deficient[9] and thus highly reactive metal-organic compound that has been applied for ALD of $AlO_x$ at low temperatures, even below 25 °C.[10] *In situ* studies of the initial growth regime during ALD of $AlO_x$ have revealed that TMA can directly react with the native oxides of III-V semiconductors, resulting in the formation of aluminum oxide *via* interconversion and ligand exchange reactions.[11] This high reactivity of TMA with the substrate has been previously applied to passivate the surfaces of III-V semiconductors *in situ* before deposition of an ALD dielectric.[11a, 11b] In later work, the sequential exposure of III-V surfaces to a hydrogen ($H_2$) plasma and TMA in an ALD process proved to be even more effective for passivation of the surface oxide layers.[12] Despite the intriguing chemical characteristics of oxidant-free ALD oxide interconversion reactions, neither the potential for creating ultrathin conformal oxides down to the monolayer limit nor the chemical functional characteristics of the resulting layers have been explored so far.

Understanding film nucleation is key for realizing ultimately thin coatings by ALD. Though the nucleation behavior is well established for thermal ALD of $AlO_x$ using TMA and water,[13] little is known about film formation in oxidant-free processes involving TMA and $H_2$ plasma.

Here, we evaluate the AlO$_x$ growth evolution on free-standing *c*-plane GaN during cyclic exposure to TMA and atomic hydrogen by *in situ* monitoring of the ALD film thickness using spectroscopic ellipsometry (SE), as well as by complementary *ex situ* analysis using atomic force microscopy (AFM) and X-ray photoelectron spectroscopy (XPS). We find a window of self-limiting oxide growth resulting from the consumption of terminal oxygen moieties of the GaN native oxide layer during AlO$_x$ formation. Once monolayer coverage is achieved, the surface is deactivated for subsequent chemisorption of TMA, resulting in self-limited growth with no island formation. The thickness of the closed layer, independently measured by SE, AFM and XPS to be ~3 Å, agrees well with the theoretically predicted thickness for a single monolayer.[14] Creation of this single AlO$_x$ monolayer leads to a significant reduction of the work function of GaN by 0.38 eV. Furthermore, relative to the bare GaN surface, the monolayer AlO$_x$ provides additional chemical functionality, which is highlighted by its high reactivity with phosphonic acids to form self-assembled organic monolayers with a coverage that approaches the theoretical limit at room temperature. Thus, the GaN/monolayer AlO$_x$ system provides a novel platform for creating self-assembled monolayers with strong electronic coupling to the underlying semiconductor due to the absence of an extended interlayer. Not only does this work provide a new approach to creating dielectric films at the monolayer limit, but it also creates opportunities for engineering electronic and chemical properties of functional interfaces of key relevance in optoelectronics, (photo)catalysis, and energy storage applications.

## 2. Results and Discussion

As a starting point for elucidating the growth of $AlO_x$ on GaN substrates *via* the oxidant-free plasma-assisted ALD process, *in situ* spectroscopic ellipsometry was used to track film thickness dynamics during the different steps of each cycle, as well as over multiple successive cycles. As shown in **Figure 1**a, which provides a plot of film thickness as a function of time, three different growth regimes can be distinguished during 100 cycles of TMA and hydrogen plasma exposure. These regimes are characterized by (i) growth with an exponentially decaying rate, (ii) a saturated film thickness with a growth rate of zero, and (iii) slow growth with an approximately constant rate. Each of these is discussed in detail below.

In regime (i), the growth-per-cycle (GPC), which is defined as the measured change in layer thickness per cycle, follows an exponential decay and reaches saturation (*i.e.*, GPC = 0 Å) after 8 cycles with a thickness of 2.8 ± 0.1 Å (Figure 1b). The origin of this saturation behavior can be understood by considering the detailed growth mechanism. As shown in Figure 1b, the first few cycles exhibit the largest GPC, which suggests a substrate-enhanced growth governed by the accessible binding sites on the GaN surface.[15] This stands in stark contrast to the growth characteristics of traditional thermal ALD of $AlO_x$ on GaN using TMA and $H_2O$, in which growth is inhibited during the initial cycles (Figure 1b,c).

XPS analysis revealed that the surface of GaN is coated with a ~1 nm thin native oxide layer that is terminated with OH groups (Supporting Information S5). The thickness of this native oxide layer was confirmed by X-ray reflectivity (XRR) (Figure S2), and agrees with the literature.[16] Consistent with XPS analysis, static water contact angle measurements of these surfaces exhibit hydrophilic character, with contact angles of ~39° and ~34° obtained before and after the initial $H_2$ plasma treatment (Table S2). This hydrophilicity indicates that the native gallium oxide surface is partially hydroxylated, thereby providing Ga-OH binding sites that can react with TMA through a ligand-exchange reaction, in a manner similar to TMA reacting with SiO$_2$.[13a] As a consequence, exposure to TMA results in the formation of AlO$_x$ according to the following ligand exchange reaction:[17]

$$Ga - OH + Al(CH_3)_3 \rightarrow Ga - O - Al(CH_3)_2 + CH_4 \qquad (1).$$

Although **Reaction 1** is self-limiting, the available binding sites cannot be completely consumed after one TMA half-cycle due to steric hindrances. In particular, binding sites are shielded by the methyl groups of neighboring chemisorbed TMA molecules and may also be blocked by physisorbed TMA molecules and clusters. While weakly bound TMA molecules are swept away from the surface during argon (Ar) purging, the ligands of chemisorbed TMA molecules will remain until chemically reactive species are introduced. As indicated by the red arrows of Figure 1b, subsequent H$_2$ plasma exposure leads to a decrease in the adsorbate thickness, indicating that atomic hydrogen, which is used here instead of an oxidant, reduces the adsorbed TMA *via* the formation of methane. Consequently, Ga-OH binding sites that were blocked by methyl ligands after the first TMA half-cycle become accessible for subsequent reactions with TMA. This simple mechanistic model predicts an exponential decay of available binding sites[18] and explains the observed exponential decrease of the individual cycle growth rate over successive cycles.

A critical aspect of this oxidant-free AlO$_x$ ALD mechanism is that the surface is chemically deactivated as the Ga-OH sites are consumed, as indicated by the observed decrease of the GPC, which reaches zero after eight TMA/hydrogen plasma cycles. Since exposure to hydrogen plasma reduces the methyl ligands of chemisorbed TMA to methane, an O-Al* terminal surface, which may include partial coverage by O-Al-CH$_3$ sites, is established. This surface composition inhibits further chemisorption of TMA and, thus, the growth of subsequent layers, thereby giving rise to the second regime, (ii), during which the thickness remains nearly constant (Figure 1a,b). In addition, the *in situ* SE measurements reveal that the AlO$_x$ film is not chemically reduced upon exposure to several cycles of low power H plasma. Here, we note that the AlO$_x$-coated GaN exhibits a water contact angle of ~8°, which indicates a strongly hydrophilic surface (Table S2) that is consistent with an Al-OH surface termination. However, these measurements require exposure to ambient atmosphere, which is expected to rapidly oxidize the -Al* termination, thereby introducing hydrophilic oxygen groups on the surface. The proposed growth mechanism resembles a Frank–van der Merwe type of growth, in which TMA reacts preferentially with substrate surface sites, resulting in complete surface coverage. However, in contrast to conventional growth modes, deposition of subsequent layers is suppressed due to chemical deactivation of the surface. Importantly, this chemical inactivity reduces the likelihood of island-formation[13b] and provides the intriguing opportunity for ideally self-saturating growth of $AlO_x$ down to the single monolayer limit. To explore this possibility, we consider the saturated $AlO_x$ film thickness and compare it to the predicted monolayer thickness, which can be estimated according to **Equation 2**,

$$h_{ml} = \left(\frac{M}{\rho N_A}\right)^{1/3} \qquad (2),$$

where $M$, $\rho$, and $N_A$ are the molar mass, density, and Avogadro constant, respectively. For an ultrathin layer of this type, it is difficult to determine the density. However, reasonable values can be obtained from prior reports of ALD alumina, which yield values in the range of 3.2 – 3.5 g/cm$^3$ at similar substrate temperatures.[19] At the level of a monolayer, half the volume of the α-$Al_2O_3$ unit cell (*i.e.*, a nominal composition of $AlO_{1.5}$) is assumed, which yields an approximate thickness, $h_{ml}$, in the range of 2.89 to 2.98 Å. Thus, the film thickness in the saturation regime (ii) of 2.8 ± 0.1 Å, determined experimentally *via in situ* spectroscopic ellipsometry, is in excellent agreement with the predicted thickness of a single monolayer of alumina. This agreement is fully consistent with the proposed film formation mechanism and offers fascinating possibilities for the establishment of two-dimensional dielectric interlayers, as discussed below.

Although no growth is observed for several subsequent TMA/H$_2$ plasma cycles, a third growth regime, indicated as the region (iii) in Figure 1a, emerges after 20 cycles. This regime is characterized by a constant, but slow, growth rate of 0.03 Å per cycle. According to the model discussed above, no such growth would be expected. However, *in situ* SE measurements suggest that repetitive exposure to $H_2$ plasma and TMA pulses leads to accumulation of material on the surface, indicating film growth at a comparably slow rate. Although the origin of this growth regime is not conclusively known, it is apparent that TMA precursor adsorption on the surface must occur, but with very low coverages, suggesting a potential role of defects as binding sites. Chemical interactions of TMA and such non-idealities may yield dissociative chemisorption,[20] and the cumulative effect of repetitive $H_2$ plasma exposure would then result in a complete reduction to elemental Al since no oxygen from the underlying oxide is accessible. To confirm the feasibility of metallic Al growth, we prepared a sample using 200 cycles of TMA/$H_2$ plasma at a higher plasma power (300 W). The larger growth rate (GPC ≈ 0.15 Å) resulted in Al films that are sufficiently thick to not completely oxidize in air. XPS analysis of the resulting surfaces revealed an Al 2p component at a lower binding energy (~72.9 eV), indicative of metallic Al, alongside the higher binding energy feature associated with Al-O binding (~75.6 eV) (Figure S10). This result, which is consistent with the prior report of metallic Al ALD using TMA/$H_2$ plasma cycles,[21] lends additional credence to the proposed description of linear growth in regime (iii).

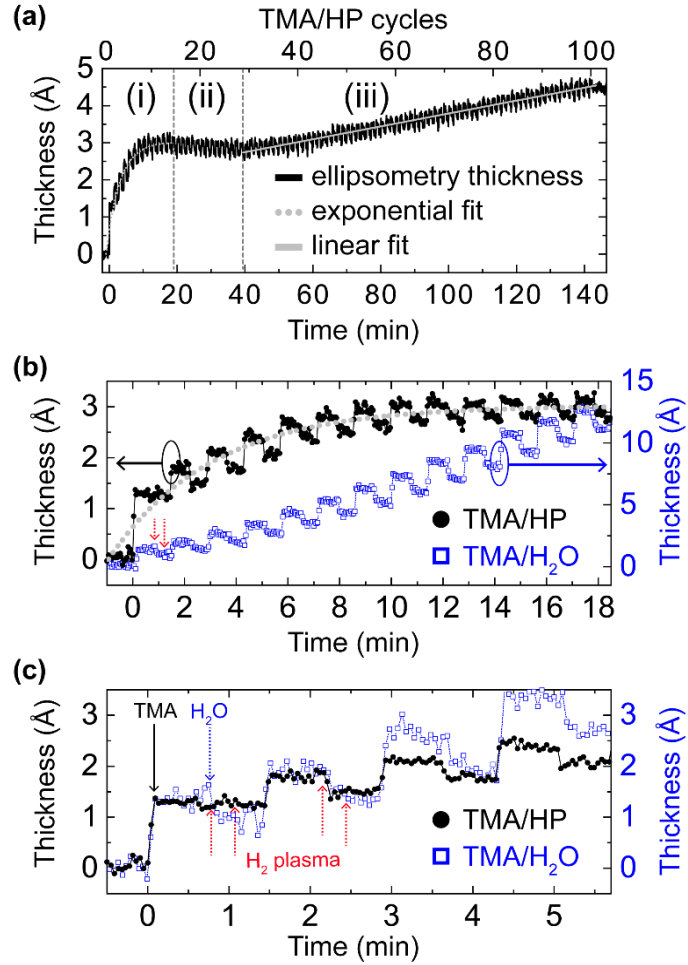

**Figure 1.** (a) Plot of the ALD layer thickness as a function of time, determined *via in situ* spectroscopic ellipsometry. Three different growth regimes (i – iii) can be distinguished. (b) Zoom-in of the thickness-time plot showing growth regime (i), in which the growth-per-cycle exponentially decreases before saturation at ~3 Å. For comparison, the thickness-time plot of a thermal ALD AlO$_x$ process with H$_2$O is shown (blue, thickness scale given by right axis). (c) Zoom-in of the thickness-time plot showing the first four cycles. The black, blue and red arrows indicate TMA injections, H$_2$O injections and plasma times, respectively, for the first two cycles. The reactor temperature was maintained at 280 °C, and free-standing *c*-plane GaN substrates were used for both recipes.

We now return to a detailed analysis of the characteristics of the ultrathin AlO$_x$ layer formed prior to the onset of regime (iii), as well as the properties of its interface with GaN. To assess ALD film conformality and surface roughness, both bare GaN and GaN/AlO$_x$ were characterized by AFM. The AFM height images before and after 20 cycles of TMA and H$_2$ plasma exposure showed a nearly identical root mean square (rms) roughness of 237 ± 6 pm and 228 ± 3 pm, respectively, indicating the formation of a uniform and continuous ALD film conforming to the topography of the GaN surface (**Figure 2**a,b). The normalized height distributions of bare and TMA/H$_2$ plasma-treated GaN follow symmetric Gaussian lineshapes (skewness = ±0.05, kurtosis = 2.95) typical of homogenous surfaces with randomly distributed height variations. No indications of pin-holes or island growth were observed, which is consistent with the proposed self-limiting growth mechanism. Following characterization of the topography, a diamond-like carbon (DLC) tip was used to abrade the AlO$_x$ monolayer in contact mode, as previously demonstrated with thicker ALD AlO$_x$ films on Si substrates.[22] The subsequently acquired AFM phase and height images suggest the presence of a continuous AlO$_x$ monolayer with a thickness of 3 ± 1 Å (Figure S14), consistent with the thickness derived from SE measurements.

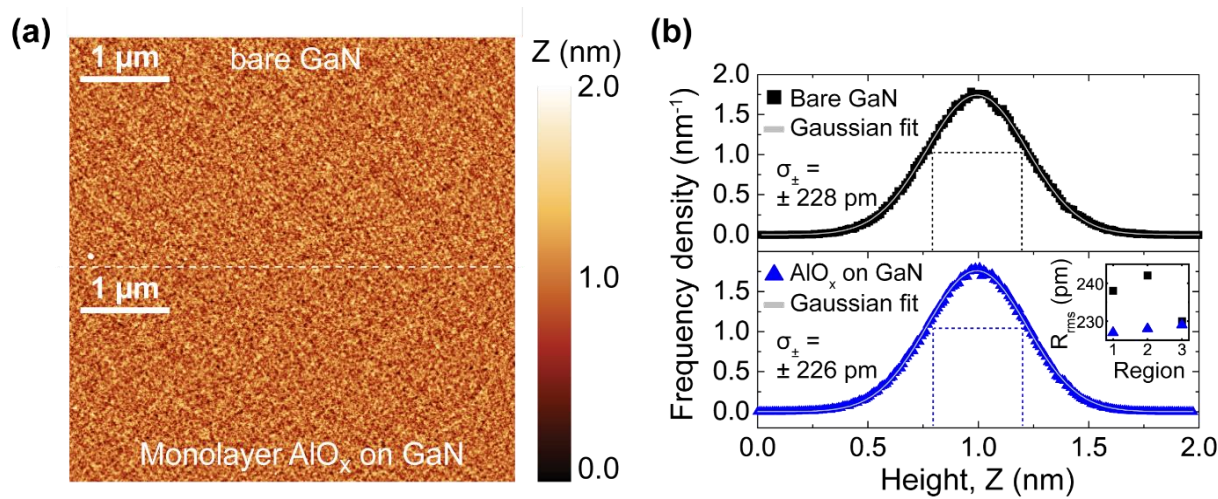

**Figure 2.** (a) AFM topographs of bare (upper image) and monolayer AlO$_x$-coated GaN (lower image) following 20 ALD cycles of H$_2$ plasma and TMA. (b) The frequency densities extracted from the AFM height images follow a Gaussian distribution with nearly identical parameters.

XPS was performed to determine the chemical changes to the surface arising from exposure of GaN to 20 cycles of TMA and H$_2$ plasma. **Figure 3**a shows the Ga 3d core level region of the bare GaN surface, within which Ga-N, Ga-O, and Ga-Ga, as well as N 2s, components can be discriminated. Spectral fitting was achieved by constraining the spin-orbit splitting to be 0.43 eV[23] with an area ratio of 3:2 and equal FWHM for both components of the 3d doublets. For both the bare and TMA/H$_2$ plasma-exposed sample, the magnitude of the shift of the Ga-O components relative to the main Ga-N photoemission is consistent with prior studies[24] and indicates the presence of a native oxide comprising primarily $Ga_2O_3$. Ga 3d photoelectrons emitted from Ga-O-Al environment are expected to shift towards lower BEs, *i.e.* closer to the Ga-N main peak because of the relatively lower electronegativity of aluminum. Indeed, XPS analysis indicates that the Ga-O/Ga-N ratio decreased after the TMA/$H_2$ plasma process (Figure 3a, inset). This finding suggests an $AlO_x$ film growth mechanism that includes the formation of Ga-O-Al bonds,[25] which is in agreement with the proposed Reaction 1 based on the real-time SE data, as well as previous reports of TMA reactivity on GaN[24, 26] and AlGaN[27] during traditional thermal ALD using $H_2O$ as an oxidant.

The Al 2p core level spectrum is well described by a single component with a binding energy of 75.40 eV, which can be assigned to an Al-O environment (Figure 3b). Comparison of the 1.71 eV FWHM to the instrumental resolution of ~0.30 eV (Figure S6), indicates significant inhomogeneous broadening that is typical of amorphous materials. By comparison, the Al 2p peak of a 25 nm thick $AlO_x$ reference film grown by thermal ALD (250 cycles of TMA/$H_2O$, 280 °C) on GaN has an FWHM of 1.43 eV (Figure S11), in agreement with prior literature reports.[28] The O 1s spectra of bare and $AlO_x$-coated GaN substrates reveal two chemical components attributed to hydroxyl groups and metal-oxide bonds (Figure S8).

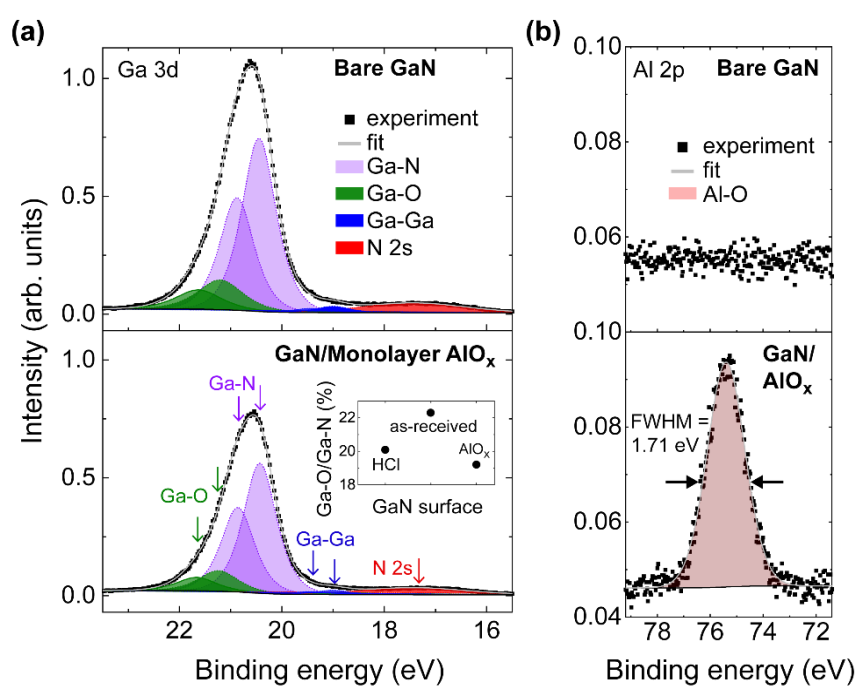

**Figure 3.** X-ray photoelectron spectra of (a) Ga 3d and (b) Al 2p core levels of bare GaN (top row) and monolayer AlO$_x$-coated GaN (bottom row). The inset in (a) shows the average gallium oxide concentrations for bare (as received and after HCl etching) and monolayer AlO$_x$-coated GaN surfaces.

It is expected that the hydroxylated monolayer AlO$_x$ coating should enhance surface basicity (acid-base reactivity), thereby providing the opportunity to introduce new chemical functionality to the surface. To test this hypothesis, we investigated the reactivity of the ALD-modified GaN surface with phosphonic acids (PAs) under standard conditions. The PA head group binding sites, namely the acidic hydroxyls and the phosphoryl group, can react with the hydroxyl groups of the substrate *via* an acid-base heterocondensation.[29] While this reaction requires thermal activation for acidic and weakly basic (Lewis acidity) hydroxylated surfaces, such as the native oxides of Ga [30] and Si [31], it can occur under standard conditions for basic hydroxylated substrates, such as aluminum oxide.[32] To test the reactivity of the ALD-modified surface, we performed an organophosphonic acid functionalization reaction with 11-hydroxyundecylphosphonic acid (PA-C11-OH) *via* a modified immersion technique at room temperature and atmospheric pressure (Supporting Information S7).

The AFM topography (**Figure 4**a,b) demonstrates a uniform and smooth organic SAM with an rms roughness of 274 ± 9 pm, thereby providing further evidence of a continuous and pin-hole free AlO$_x$ coating. An average organic SAM thickness of ~5.8 ± 0.4 Å was determined from the step heights after removal of the organic layer in contact mode AFM (Figure 4a,b). Considering the length of PA-C11-OH and the measured SAM thickness, a molecular tilt angle, $\theta$, of 70.5 ± 2.8° with respect to the surface normal was determined. This value is relatively large compared to prior literature reports of PA-C11-OH SAMs. For example, (100) Si yields tilt angles of ~45°.[33] However, the tilt angle and thus the SAM thickness are related to the PA binding configuration, which is known to be affected by the deposition method, PA structure, and surface coverage.[34] In particular, the PA can interact with the underlying AlO$_x$ layer *via* P–O–Al bonding in mono-, bi-, or tridentate modes, depending on the number of PA functional groups that are involved, as indicated in Figure 4d.[29] Previously, bidentate bonding was reported to be dominant for GaN following functionalization at elevated temperature,[35] while a range of different binding motifs has been reported for aluminum oxide.[34]

To probe the binding motif and quantify the molecular coverage, XPS measurements were performed on functionalized surfaces. Figure 4c shows a typical XPS spectrum of the P 2p core level region, which also includes the Ga 3p plasmon, for a GaN/monolayer AlO$_x$/PA-C11-OH sample. Despite a theoretical spin-orbit splitting of 0.9 eV,[36] the main P 2p peak at 135 eV is broad (FWHM = 1.79 eV) and approximately symmetric, which is consistent with a mixture of binding modes.[36-37] The optimum fit of the P 2p spectrum (Figure 4c, red line) suggests a combination of two binding configurations, of which the mono-dentate binding motif is dominant (88 ± 3 %).

Prior reports attributed the evolution from higher to lower denticities to an increasing SAM surface coverage.[37-38] Our results are consistent with this finding since we observe a relatively large phosphorus atom density of $(4.5 \pm 0.3) \times 10^{14}$ cm$^{-2}$, corresponding to a PA surface coverage, $n$, of 4.5 ± 0.3 nm$^{-2}$, as calculated from the XPS data (Figure 4c) according to the formalism introduced by Kim *et al.*[30] (Supporting Information S7). This high SAM surface coverage (4.5 ± 0.3 nm$^{-2}$) on GaN/monolayer AlO$_x$ approaches the theoretical limits predicted for aluminum oxide (4.65 nm$^{-2}$)[39] by DFT and estimated from the molar volume of phosphorus acid (4.25 nm$^{-2}$),[40] and it is two times higher than reported for alkyl-PAs grafted on *c*-plane-Ga-polar GaN (2.3 nm$^{-2}$).[30] These high coverages are consistent with the formation of a continuous AlO$_x$ coating, onto which the PAs molecules assume a predominantly mono-dentate binding motif in a high-density SAM.

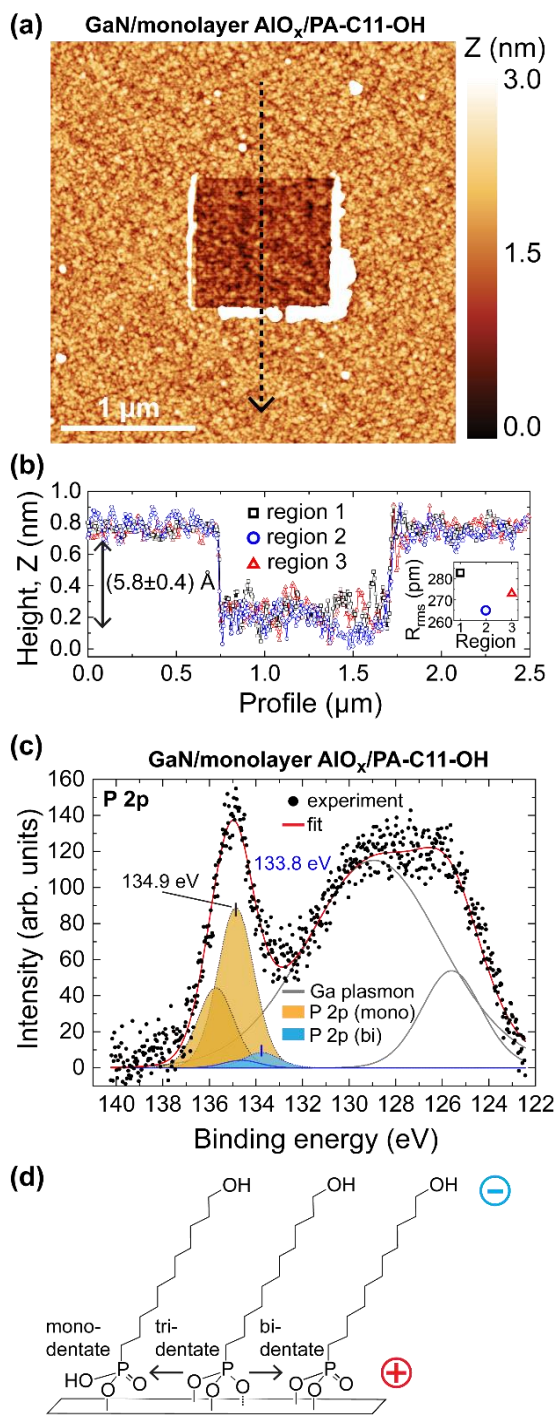

**Figure 4.** (a) AFM topography of monolayer AlO$_x$-coated GaN functionalized with PA-C11-OH showing a center region where the organic layer was removed by contact mode AFM. (b) Height profiles collected in three different regions, each located a few mm apart from one another, indicating the homogeneity of the SAM layer and allowing determination of the thickness. (c) P 2p photoelectron spectrum and corresponding fit of the GaN/monolayer AlO$_x$/PA-C11-OH sample. P 2p spin-orbit splitting and Ga plasmons are considered in the fit, which suggests a mostly monodentate binding motif (P 2p (mono)) of the PA head group to the monolayer AlO$_x$. (d) Schematic representation of three possible binding configurations of PA-C11-OH on a hydroxylated surface.

Having established the surface chemistry of the GaN/monolayer AlO$_x$ system, we next focus on the effect of the monolayer AlO$_x$ coating on the surface energetics, which can influence electrical transport, surface chemical reactions, and dipole interactions. To quantify how the deposition of monolayer AlO$_x$ and SAM functionalization affect the work function, $\Phi$, and surface band bending of GaN, we performed contact potential difference (CPD) and surface photovoltage (SPV) measurements in a vacuum (**Figure 5**a). After deposition of monolayer AlO$_x$ on GaN, CPD measurements indicate that the work function decreases by 0.38 eV compared to bare GaN. Concurrently, SPV measurements reveal an upward band bending of 0.49 eV, which is the same for both bare GaN and monolayer AlO$_x$/GaN. This is consistent with XPS measurements, which are sensitive to changes of band bending but not work function: here, XPS revealed that the VB edge positions (Figure S12) and core level BEs (Table S4) were not affected by the presence of the AlO$_x$ monolayer to within the measurement error. Thus, changes to the interfacial energetics can be explained in terms of a change of the surface dipole associated with monolayer AlO$_x$ deposition (Figure 5b). The polarity of this dipole, for which the positive endpoints outwards, is consistent with the smaller electronegativity of Al (1.7 [41]) relative to Ga (2.4 [41]). In addition, polarization-induced negative bound charges of the Ga-polar GaN may facilitate interface dipole formation as these surface charges are compensated by the native GaO$_x$ layer,[42] rendering the GaO$_x$ at the GaN/GaO$_x$ interface more positive compared to the GaO$_x$ surface. The microscopic dipole that forms at the interface between the native GaO$_x$ film and the sub-stoichiometric monolayer AlO$_x$ results in an abrupt change of the vacuum level that reduces the effective work function (Figure 5b). The bound charge density, $\sigma$, at the monolayer AlO$_x$/GaO$_x$ interface can be estimated using Gauss' theorem, $\sigma = \frac{|\Delta V_{\text{AlOx}}|\varepsilon_{\text{eff}}\varepsilon_0}{de_0}$, where $|\Delta V_{\text{AlOx}}|$ is the magnitude of the measured CPD change after monolayer AlO$_x$ deposition, $\varepsilon_{\text{eff}} = \frac{2\varepsilon_{\text{AlOx}}\varepsilon_{\text{GaOx}}}{\varepsilon_{\text{AlOx}}+\varepsilon_{\text{GaOx}}}$ is the effective dielectric constant, $d$ is the separation between positive and negative point charges of the dipole layer, and $e_0$ is the elementary charge.

Assuming $d = 3$ Å as the average distance between Ga and Al atoms at the Ga-O-Al interface, $\varepsilon_{AlOx} \sim 4$ for ultrathin ALD AlO$_x$ [18] and $\varepsilon_{GaOx} \sim 9$ for amorphous GaO$_x$ [43] yields $\sigma = 3.5 \times 10^{13}$ cm$^{-2}$, which is in the range of polarization-induced charge densities of polar GaN. We note that in addition to the interface dipole, fixed charges of both polarities have been measured for thicker AlO$_x$ films by resonant XPS,[44] predicted by density functional theory (DFT) calculations,[45] and quantified by capacitance-voltage measurements, respectively.[46] [47] However, for thicker oxide layers, fixed charges would create an extended electric field and further influence the work function.

CPD measurements were also performed to determine how the PA-C11-OH SAM affects the surface energetics relative to the bare GaN and GaN/AlO$_x$ surfaces, as shown in Figure 5a. The measured increase in the CPD after GaN/monolayer AlO$_x$ surface functionalization with a PA-C11-OH organic monolayer is consistent with the presence of a surface dipole that arises from the permanent OH surface termination introduced by the SAM (Figure 5c). The surface dipole introduced by the AlO$_x$ monolayer is in the opposite direction to the SAM dipole and therefore may account for the large tilt angle of the organic layer. The out-of-plane component of the dipole moment, $\mu_z$, of an adsorbed monolayer can be calculated using the Helmholtz relationship, $\mu_z = \frac{\Delta V_{PA} \varepsilon_{PA} \varepsilon_0}{n \cos \theta}$,[48] where $\varepsilon_{PA}$ is the relative dielectric constant, $\varepsilon_0$ is the vacuum permittivity, and $n$ is the areal density of adsorbed molecules (dipoles). Inserting the surface coverage of $4.5 \times 10^{14}$ cm$^{-2}$ derived from XPS analysis, the measured change of the CPD, $\Delta V_{PA}$, of 0.18 V, and the reported alkyl SAM permittivity, $\varepsilon_{PA}$, of 3,[49] yields a surface dipole of +0.32 D. By correcting for the tilt angle, $\theta \sim 71°$ deduced from the measured SAM thickness (Figure 5b), the surface dipole of a vertically aligned PA-C11-OH monolayer would equal 0.95 D. While this value is lower than the reported free molecule dipole moment (+1.46 D) predicted by density functional theory,[50] it has been established that depolarization within SAMs leads to a deviation between the free molecular dipole and the bound molecular dipole, of which the latter was found to correlate with observed changes in the work function.[51] Here, we find that surface energetics of the GaN are significantly modified by the introduction of a single monolayer of AlOx and can be additionally tuned by the introduction of a molecular monolayer, whose assembly is enabled by the chemical functionality of the ALD oxide.

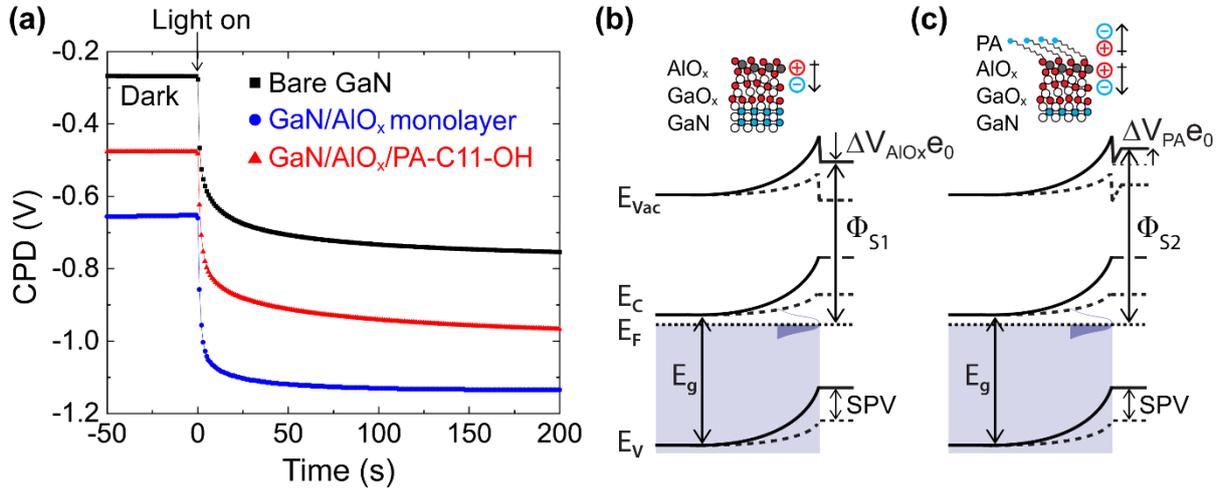

**Figure 5.** (a) CPD evolution for the three sample systems: bare GaN (black squares), monolayer AlOx-coated GaN (blue circles), and GaN/monolayer AlOx/PA-C11-OH (red triangles). The SPV is measured with above-bandgap illumination (340 nm wavelength). (b),(c) Schematic band diagrams and illustrations of the GaN/GaOx/monolayer AlOx and GaN/GaOx/monolayer AlOx/PA-C11-OH samples, respectively, in which the + and – indicate the dipole orientations in the different layers.

## 3. Conclusion

The presented oxidant-free ALD process allows for the fabrication of conformal and pin-hole free oxide coatings down to the single monolayer limit. This was demonstrated *via* the deposition of a monolayer AlO$_x$ film on GaN, which was accomplished by oxidant-free ALD with TMA and H$_2$ plasma. The TMA/H$_2$ plasma growth process self-terminates after complete surface coverage with AlO$_x$ because the TMA preferentially reacts with the surface sites on the gallium oxide, but not the formed AlO$_x$ layer, thereby preventing vertical (three-dimensional) growth. The average thickness of the sub-nanometer thin AlO$_x$ coating on the GaN surfaces was determined with statistical confidence by SE and XPS and is consistent with the predicted monolayer AlO$_x$ thickness. The ultrathin AlO$_x$ coating increases surface hydrophilicity and basicity, facilitating the formation of a stable SAM under standard conditions with a coverage approaching the theoretical limit, thereby allowing additional tuning of interfacial energetics and the controlled introduction of chemical moieties at the terminal surface.

This work demonstrates, to the best of our knowledge, the first experimental realization of monolayer AlO$_x$ on a non-metal surface. Given the shared surface reactivity, the presented monolayer AlO$_x$ scheme based on oxidant-free plasma-assisted ALD can be extended to the broader range of III-V semiconductors for the formation of different monolayer coatings derived from ALD precursors possessing strong reduction potentials (*e.g.*, tetrakisdimethylamido-hafnium). Importantly, this approach overcomes a major non-ideality of traditional ALD by eliminating island formation during the growth. Thus, it offers a powerful approach for realizing precise heterostructures by fulfilling the rigorous definition of *atomic layer* deposition. The resulting control of surface chemistry and energetics at the single monolayer level provides functional control of surfaces tailored for use in chemical sensing, (photo)catalysis, high electron mobility transistors, and advanced optoelectronic devices.

## 4. Methods

### 4.1 GaN substrates

Free-standing and unintentionally silicon (Si) doped (resistivity < 0.5 Ω cm; donor density ≤ $1\times10^{17}$ cm$^{-3}$) $c$-plane Ga-polar GaN substrates (10×10.5 mm$^2$), grown by hydride vapor phase epitaxy (HVPE) by the Nanowin Science and Technology company (Suzhou, China) were used for the ALD nucleation study, as well as for surface functionalization and chemical analysis by XPS. The GaN substrates have a nominal dislocation density below $1\times10^{6}$ cm$^{-2}$ and were chemomechanically polished (epi-ready) by the manufacturer. Prior to characterization and ALD processing, the GaN substrates were sonicated (37 kHz, 120 W) successively in acetone, isopropanol, and deionized water for 10 minutes, respectively, and blow-dried by nitrogen. X-ray diffraction (XRD) was performed with a Rigaku SmartLab diffractometer equipped with a Cu anode and a 2× Ge(220) monochromator. XRD analysis confirmed that the substrate surface is parallel to the $c$-plane of GaN (Figure S1). Diffraction peaks from the (20-24) plane are separated by 60°, consistent with the six-fold symmetry of the hexagonal structure of wurtzite GaN. The full width at half maximum (FWHM) of the x-ray rocking curve for the (0002) plane is 87 arcsec, verifying the high crystal quality of the GaN. Templated $c$-plane GaN layers on sapphire were used for static water contact angle measurements (Supporting Information S4). Single side polished 20 µm thick $c$-plane, Ga-polar GaN templates (Si-doped, carrier concentration >$1\times10^{17}$, dislocation density < $5\times10^{8}$ cm$^{-2}$) grown by HVPE on 430 ± 25 µm thick sapphire substrates were obtained from MSE supplies LLC (Arizona, USA). The individual GaN wafers were cut into 10×10 mm$^2$ pieces.

### 4.2 Atomic layer deposition

GaN substrates were coated with AlO$_x$ in a hot-wall plasma-enhanced atomic layer deposition (PE-ALD) reactor (Fiji G2, Veeco CNT) in continuous flow mode. An *in situ* H$_2$ plasma pretreatment (2 cycles, 3 sec, 100 W, 0.02 Torr) at 280 °C was performed to remove the approximately 5 Å thin adsorbate layer of air-exposed GaN (Figure S4) prior to deposition of alumina. Conveniently, hydrogen treatment can also passivate dangling bonds on GaN.[52] $AlO_x$ films were grown using TMA (electronic grade, 99.999 %, STREM Chemicals) as the precursor and Ar (99.9999 %, Linde) as the carrier gas during the first half-cycle at ~0.09 Torr while the turbo pump was isolated. The pressure-time plot (Figure S5) illustrates the ALD $AlO_x$ growth process, which has previously been implemented with similar parameters for the *in situ* preparation and passivation of III-V semiconductor surfaces before deposition of dielectrics.[12a, 53] The precursor and plasma doses were set such that the reactions with the GaN surface were self-limiting, which was determined by *in situ* SE monitoring of the adsorbate thickness during plasma and TMA dose tests. The reactor wall and chuck temperatures were controlled to 280 °C, *i.e.*, below the TMA decomposition temperature (≥ 300 °C). During the second half-cycle, $H_2$ (99.9999 %, Linde) was supplied as the carrier and plasma gas at a base pressure of 0.02 Torr, which was achieved with a turbo pump. $H_2$ plasma was generated in a sapphire tube with an inductively coupled plasma source, where a radio frequency (RF) bias at 13.64 MHz and 100 W was applied to a copper coil wrapped around the sapphire tube. Each cycle of ALD $AlO_x$ followed the sequence: 0.08 s TMA dose, 30 s Ar purge, 10 s $H_2$ purge, 2 s $H_2$ plasma, 10 s Ar purge, 10 s $H_2$ purge, 2 s $H_2$ plasma, 10 s Ar purge.

**4.3 Preparation of self-assembled monolayers of phosphonic acids (SAMPs)**

SAMPs of 11-hydroxyundecylphosphonic acid (≥ 99 % (GC); referred to as PA-C11-OH) (SiKÉMIA, Montpellier, France) were prepared *via* a modified immersion technique, which is described in detail in section S7 of the Supporting Information.

**4.4 Spectroscopic ellipsometry**

Changes in the GaN surface oxide layer thickness were monitored in real-time using an *in situ* ellipsometer (M-2000, J. A. Woollam) with a sampling time of ~ 3 s during ALD. The dwell (purge) times after each precursor and plasma step were programmed to be at least three times the SE integration time. A general oscillator model was used to model changes of the polarization in the wavelength range around the GaN band gap, *i.e.*, between 210 nm and 400 nm (Figure S3).

**4.5 X-ray photoelectron spectroscopy (XPS)**

XPS spectra were acquired at pass energy of 10 eV with a Kratos Axis Ultra setup equipped with a monochromatic Al Kα X-ray source. Charge neutralization was not required as no binding energy shifts indicative of (differential) charging were observed for the free-standing not intentionally doped *n*-type GaN samples. The instrumental broadening (0.30 eV) was determined by fitting of the measured Ag 3d core level spectrum of a silver calibration sample with a Voigt function (Figure S6). To facilitate the quantitative analysis of compositions and film thicknesses, the surface adsorbate layer (water and adventitious carbon) of the ambient-exposed GaN samples was selectively removed by mild sputtering with $Ar_{1000}^+$-ion clusters (10 keV, 60 s, 37 ° incidence angle) generated with a gas cluster ion source (GCIS). Comparison of GaN core level spectra before and after sputter cleaning demonstrates the attenuation effect of the carbon overlayer on GaN core level emission intensities (Figure S7). The mild *in situ* Ar-ion cluster sputter procedure was optimized to selectively remove the surface adsorbates without altering the chemical bonds in the GaN, as confirmed by the identical Ga 2p and Ga 3d core level peak shapes before and after sputtering.

**4.6 Atomic force microscopy (AFM)**

AFM measurements were carried out with a Bruker Multimode V microscope (Billerica, MA, USA) in ambient using NSG30 AFM probes (TipsNano) with a nominal tip radius of 8 nm, typical resonance frequency of 320 kHz and force constant of 40 N/m. Height images (5×5 µm$^2$ and 3×3 µm$^2$) were acquired at a scan rate of 0.5 Hz with 512-point sampling. A diamond-like carbon probe (Tap190DLC, BudgetSensors) with a nominal radius of 15 nm was used to locally remove the $AlO_x$ monolayer and self-assembled monolayers in contact mode with tip-sample forces of 3.25 µN and 0.65 µN, respectively. The tip-sample force, $F_{ts}$, in hard contact with the surface, was calculated using Hooke's law, $F_{ts} = ks$, where the measured deflection, *s*, was determined from a force-distance curve and the spring constant, *k*, was obtained from a thermal tune. Height images (5×5 µm$^2$ and 3×3 µm$^2$) were acquired at a scan rate of 0.5 Hz with 512-point sampling.

**4.7 Kelvin probe (KP) and Surface Photovoltage (SPV) characterization**

SPV measurements were carried out in vacuum (≤ 1×10$^{-5}$ bar) at room temperature using a custom-built setup equipped with a commercial Kelvin probe and controller (Kelvin Probe S and Kelvin Control 07, Besocke DeltaPhi) and a focused light-emitting diode with a wavelength of 340 nm (M340L4, Thorlabs) and intensity of ~ 35 mW/cm$^2$. A piezoelectrically driven gold grid with a diameter of 3 mm and a work function of ~4.9 eV was used as the reference electrode. The time resolution of the CPD measurements was 1.0 s.

**Supporting Information**

Additional information on the GaN substrate properties and surface preparation methods. X-ray diffraction analysis data. Static water contact angle measurements. Spectroscopic ellipsometry model. Supporting XPS and AFM data. Description of SAMs preparation and analysis.

**Author Contributions**

[⊥]These authors contributed equally. All authors have given approval to the final version of the manuscript.

**Acknowledgments**

This work was supported by the Deutsche Forschungsgemeinschaft (DFG, German Research Foundation) under Germany´s Excellence Strategy – EXC 2089/1 – 390776260 and by the DFG through the TUM International Graduate School of Science and Engineering (IGSSE). AH acknowledges funding from the European Union's Horizon 2020 research and innovation programme under the Marie Skłodowska-Curie grant agreement No 841556. OB and IDS also acknowledge support by the Federal Ministry of Education and Research (BMBF, Germany) project number 033RC021B within the CO$_2$-WIN initiative.

Supporting Information

# Aluminum Oxide at the Monolayer Limit *via* Oxidant-free Plasma-Assisted Atomic Layer Deposition on GaN

*Alex Henning[§, ⊥, *], Johannes D. Bartl[§, #, ⊥], Andreas Zeidler[§], Simon Qian[‡], Oliver Bienek[§], Chang-Ming Jiang[§], Claudia Paulus[§], Bernhard Rieger[#], Martin Stutzmann[§], Ian D. Sharp[§, *]*

[§]Walter Schottky Institute and Physics Department, Technical University of Munich, 85748 Garching, Germany
[#]WACKER-Chair of Macromolecular Chemistry, Catalysis Research Center, Technical University of Munich, 85748 Garching, Germany
[‡]Department of Chemistry, Technical University of Munich, 85748 Garching, Germany

[⊥] **These authors contributed equally.**
[*] E-mail: sharp@wsi.tum.de, alex.henning@wsi.tum.de

**Table of Contents**

## S1. X-ray diffraction analysis of free-standing *c*-plane, Ga-polar GaN

Free-standing and unintentionally silicon (Si) doped (resistivity $< 0.5$ Ω×cm; donor densty $\leq 1\times10^{17}$ cm$^{-3}$) *c*-plane Ga-polar GaN substrates (10×10.5 mm$^2$), grown by hydride vapor phase epitaxy (HVPE) by the Nanowin Science and Technology company (Suzhou, China) were used for the ALD nucleation study, as well as for surface functionalization and chemical analysis by XPS. The GaN substrates have a nominal dislocation density below $1\times10^6$ cm$^{-2}$ and were chemomechanically polished (epi-ready) by the manufacturer. The X-ray diffraction (XRD) data (**Figure S1**) demonstrates the high crystal quality of the GaN substrates. Templated *c*-plane GaN layers on sapphire were used for static water contact angle measurements (Supporting Information S4). Single side polished 20 µm nominally thick *c*-plane, Ga-polar GaN templates (Si-doped, carrier concentration $>1\cdot10^{17}$, dislocation density $< 5\times10^8$ cm$^{-2}$) grown by HVPE on 430 ± 25 µm thick sapphire substrates were obtained from MSE supplies LLC (Arizona, USA). The individual GaN wafers were cut into 10×10 mm$^2$ pieces.

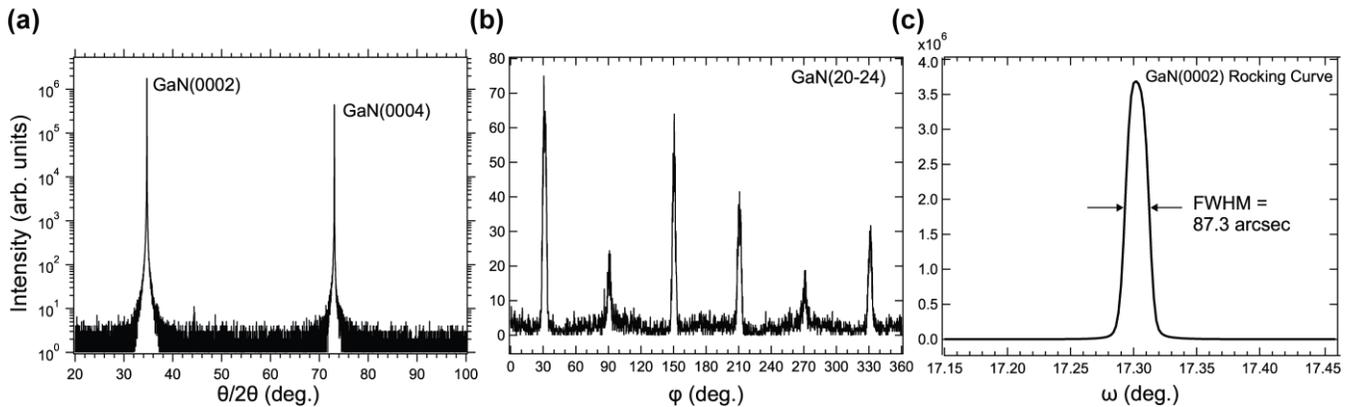

**Figure S1**. X-ray diffraction characterization of free-standing *c*-plane GaN. Diffraction peaks from (a) the (0002) and (0004) planes and from (b) the (20-24) plane. (c) Rocking curve for the (0002) plane peak.

**S2. X-ray reflectivity measurements of free-standing *c*-plane, Ga-polar GaN**

X-ray reflectivity (XRR) measurements were performed on a commercially available diffractometer (Rigaku SmartLab, Tokio, Japan) equipped with a Cu anode ($K_\alpha$ = 1.5406 Å and $K_\beta$ = 1.3922 Å) without a monochromator and a HyPix-3000 high energy resolution 2D HPAD detector in a ($\theta$-$2\theta$) geometry. The reflected intensity was recorded as a function of the momentum transfer $q$ along the surface normal (scattering vector) and normalized to 1. The scattering vector is related to the scattering angle $\theta$ via $q = \frac{4\pi}{\lambda} \sin\frac{\theta}{2}$, where $\lambda$ is the wavelength of the source. Since the $q$-range extends up to ≈ 0.65 Å$^{-1}$, the scattering length density (SLD) distribution normal to the surface can be decomposed in layers with a resolution of ≈ 4 Å according to the Fourier sampling theory.[1] The experimental data was simulated with the Python-based reflectometry analysis package refnx (ver. 0.1.8).[2] SLD profiles were converted to theoretical reflectivity curves using Abeles formalism and a differential evolution algorithm. Furthermore, Bayesian Markov-chain Monte Carlo sampling was used to estimate uncertainties of all fitting parameters (thickness, SLD, and roughness).

**Figure S2**a shows the normalized X-ray reflectivity curve (gray squares) as a function of the momentum transfer for a representative bare GaN sample. Total reflection beneath a critical momentum transfer $q_c$ of 0.046 Å$^{-1}$ is observed, which is in excellent agreement with values measured on MBE-grown GaN.[3] The inset of Figure S2a reveals periodically positioned maxima due to interference of waves reflected from the GaN/GaO$_x$ interface, where a finite discontinuity of the electron density occurs. Best fits (solid lines) were achieved by a three-layer model, considering GaN, GaO$_x$, and an adsorbate overlayer. The corresponding depth-resolved, three-layer SLD profile is presented in Figure S2b, and the obtained fitting results are summarized in **Table S1**. The average thickness of the native gallium oxide (GaO$_x$) layer is 11.3 ± 1.5 Å, and the corresponding SLD is 48.5 ± 1.5 10$^{-6}$ Å$^{-2}$ (corresponding to an electron density of 1.72 ± 0.05 e Å$^{-3}$), which agrees with reported values.[4] The electron density (0.25 ± 0.04) of the adsorbate layer is typical for hydrocarbon chains and suggests the presence of adventitious carbon on the GaO$_x$ surface,[5] which is observed in the X-ray photoelectron spectra prior to *in situ* cleaning (see Section 4).

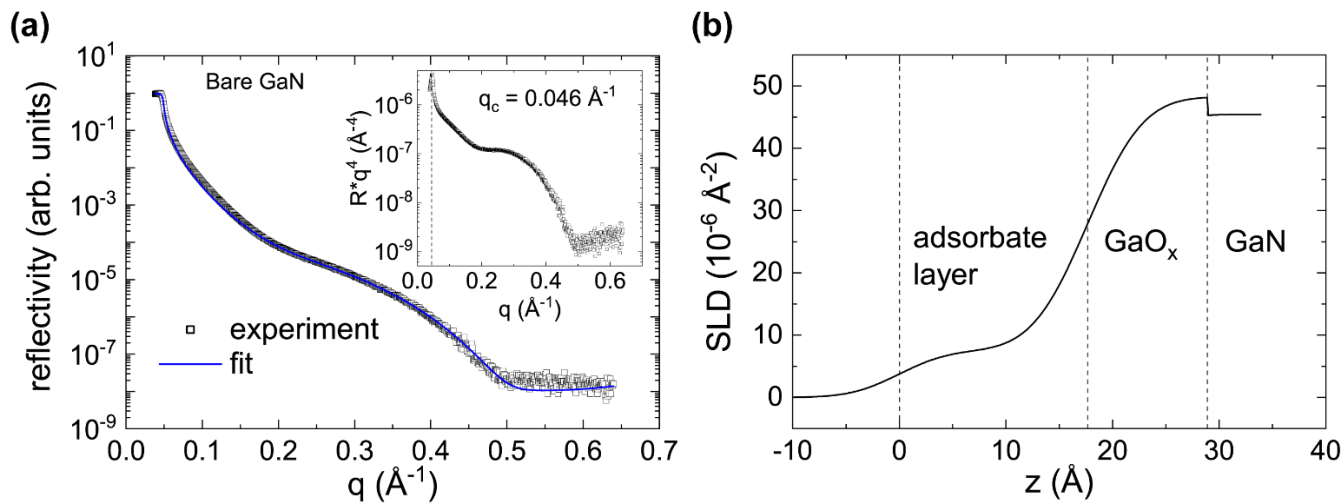

**Figure S2.** (a) Normalized X-ray reflectivity data (gray squares) as a function of the momentum transfer for a representative bare GaN sample. The best fits (solid lines) were achieved by considering a three-layer SLD model. The inset shows a superposition of the reflected intensities divided by the Fresnel reflectivity $q^{-4}$ to reveal the periodicity of the superimposed oscillating signal. (b) SLD depth profiles used to calculate the simulated intensities.

**Table S1.** Parameters obtained from fitting of X-ray reflectivity data of a representative bare GaN sample to a three-layer SLD model.

| Layer | thickness (Å) | SLD ($10^{-6}$ Å$^{-2}$) | roughness (Å) |
| --- | --- | --- | --- |
| Adsorption layer | 17.9 ± 1.1 | 7.0 ± 1.1 | 3.4 ± 0.8 |
| GaO$_x$ | 11.3 ± 1.5 | 48.5 ± 1.5 | 4.1 ± 0.2 |
| GaN | | 46.4 ± 1.5 | |
| $\chi^2$ | 21.3 | | |

## S3. *In situ* thickness measurement by spectroscopic ellipsometry during ALD

A generic oscillator model was used to model the ellipsometric polarization change in the wavelength range between 210 nm to 400 nm (**Figure S3**). The bandgap of GaN decreased to ~3.3 eV at the process temperature of 280 °C, in agreement with the literature.[6] Interferences are less pronounced in free-standing GaN compared to GaN grown on a template such as sapphire. However, backside reflections from the GaN/chuck interface are present at energies below the band gap. Therefore, the absorbing region with wavelengths near and above the GaN band gap was considered for SE data modeling. The measured thickness of native gallium oxide (11 Å) by XRR and XPS is taken into account in the SE model.

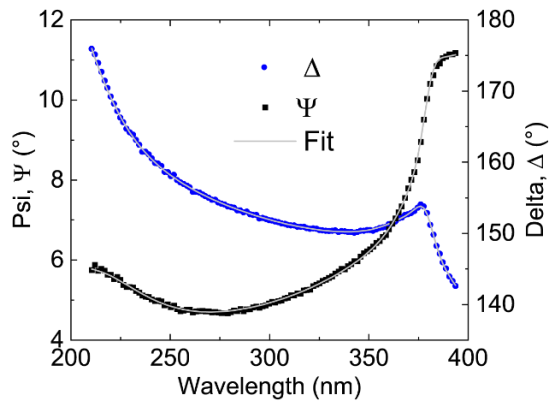

**Figure S3.** Spectroscopic ellipsometry measurement of Psi (black squares) and Delta (blue circles), and corresponding fits (grey lines) to a general oscillator model for wurtzite GaN with a layer of native oxide.

## S4. GaN surface preparation by plasma-enhanced atomic layer deposition

The *c*-plane, Ga-polar GaN substrates were exposed to two cycles of remote plasma-generated atomic hydrogen (100 W, 3 sec) prior to AlO$_x$ growth in a plasma-enhanced atomic layer deposition reactor. The overlayer film thickness decreased by ~4 Å after H$_2$ plasma exposure (**Figure S4**). The thickness after overlayer removal was defined as the starting point prior to ALD. Notably, the surface adsorbate thickness (at 280 °C) determined by SE, agrees with the carbon overlayer thickness of ~7 Å estimated from XPS analysis of GaN before and after Ar-ion cluster sputtering (Section S5). A smaller thickness was determined by *in situ* SE, likely because of partial desorption of the adsorbate layer at 280 °C. Our results suggest that the H$_2$ plasma facilitates the removal of carbon contamination on the GaN surface, previously demonstrated with forming gas for AlGaN.[7]

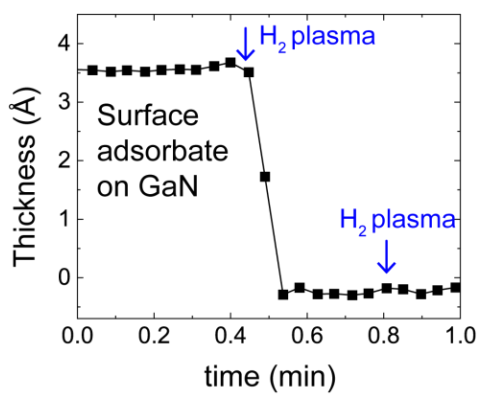

**Figure S4.** *In situ* SE thickness measurements during H$_2$ plasma treatment of an air-exposed GaN surface. The adsorbate thickness decreased after H$_2$ plasma pretreatment.

The pressure-time plot (**Figure S5**) illustrates one cycle of the monolayer AlO$_x$ growth process.

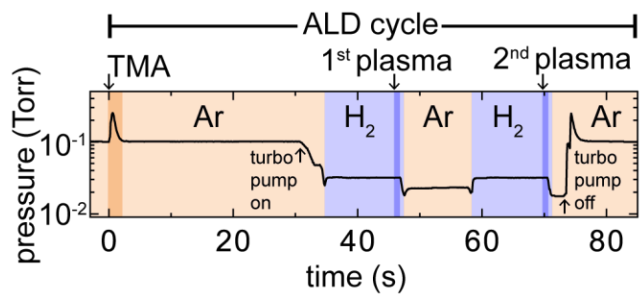

**Figure S5.** Schematic illustration of one TMA/H$_2$ plasma process cycle overlaid with the pressure-time plot recorded during AlO$_x$ growth.

Static contact angle (SCA) measurements were performed with the contact angle system OCA 15Pro (DataPhysics Instruments GmbH, Baden-Wuerttemberg, Germany) on templated *c*-plane, Ga-polar substrates under ambient conditions (27.3 °C, 32.7 % relative humidity). Data were acquired and evaluated with the basic module SCA 20 - contact angle (DataPhysics Instruments GmbH, Baden-Wuerttemberg, Germany, ver. 2.0). To estimate an average Young-LaPlace contact angle, 1 µl of deionized $H_2O$ (18.2 MΩ·cm at 25 °C, Merck Millipore) was dispensed with a rate of 1 µl/s from a high-precision syringe (Hamilton, DS 500/GT, gas-tight, 500 µl) on the sample surface and after ~3 s (reaching equilibrium) the side profile of the droplet was taken for further processing. SCAs from at least three different spots were determined to calculate a standard deviation (**Table S2**). Compared to the bare GaN substrate, the hydrophilicity of the monolayer $AlO_x$-coated surface is increased by a factor of ~ 5 (Table S2). The OH density is a crucial parameter, *e.g.* for silanization reactions,[8] but less critical for the PA surface chemistry[9] of this work. Notably, the relatively small variation (7.8 ± 0.5°) indicates that the surface is uniform after monolayer $AlO_x$ deposition and that the surface hydrophilicity is persistent in air.

**Table S2.** Static water contact angle measurement on *c*-plane, Ga-polar GaN (templates on sapphire substrates) before and after different treatments in a plasma-enhanced ALD reactor.

| treatment | process details | water CA / ° | photograph |
| --- | --- | --- | --- |
| solvent cleaning | (acetone/isopropanol in ultrasonic bath) | 52 ± 1 | 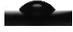 |
| annealing | 10 min at 280 °C in Ar atmosphere at 0.1 Torr | 39 ± 4 | 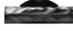 |
| $H_2$ plasma | 3 cycles $H_2$ plasma (100 W, 3 s) at 280 °C and 0.02 Torr | 33.6 ± 0.4 | 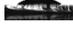 |
| TMA/$H_2$ plasma | 20 cycles TMA/$H_2$ plasma (100 W, 2 s) at 280 °C | 7.8 ± 0.5 | 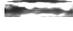 |

## S5. X-ray photoelectron spectroscopy of bare and AlO$_x$-coated $c$-plane, Ga-polar GaN

The energy resolution of the XPS system was determined from the fitting of the Ag 3d$_{5/2}$ spectrum of a silver calibration sample using a Voigt profile and yielded a value of 0.30 eV (**Figure S6**). A natural linewidth of 0.33 eV for Ag 3d$_{5/2}$ was used to deconvolute the contribution from instrumental broadening.

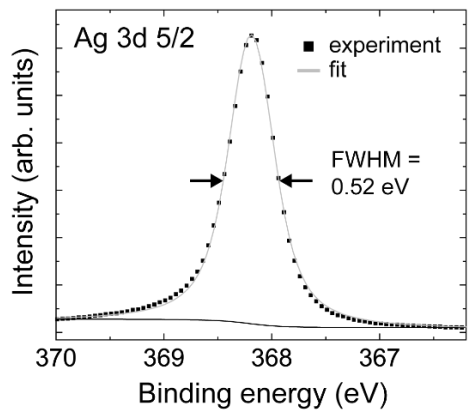

**Figure S6.** Energy calibration of the XPS system with a sputter-cleaned silver sample. An instrumental broadening of 0.30 eV was determined by fitting with a Voigt function using a natural linewidth of 0.33 eV for Ag 3d$_{5/2}$.

The substrates were cleaned by Ar$_{1000}^+$-ion cluster sputtering (10 keV, 60 s) in the analysis chamber of the XPS setup to remove surface adsorbates, which presumably comprised physisorbed water and adventitious carbon from the air-exposed GaN samples prior to XPS analysis (**Figure S7**). A BE energy shift of 0.2 eV towards lower energies is observed for the entire spectrum, indicating charging by the carbon overlayer. For larger sputtering energies ($\geq$ 20 keV), we observed removal of the native gallium oxide (not shown here). However, preferential sputtering of oxygen and nitrogen resulted in a metallic Ga-terminated surface as indicated by measured changes in Ga 3d and Ga 2p core level shapes and binding energies ($>$ 1 eV), as well as enhanced photoemission from occupied band gap states at the Fermi level in the valence band spectrum. However, no such changes were observed with more mild sputtering at 10 keV, which was used for all subsequent pre-analysis surface preparations.

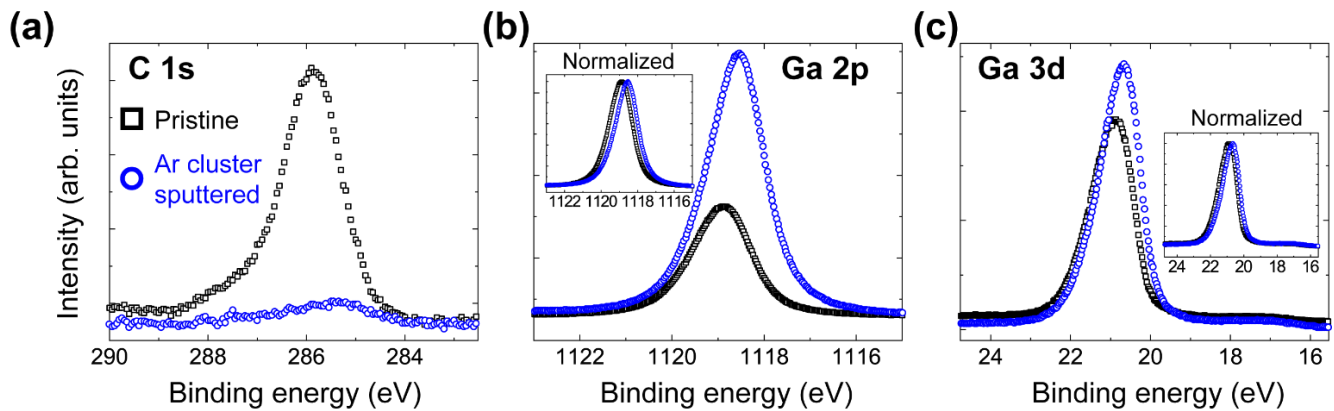

**Figure S7.** XPS of (a) C 1s, (b) Ga 2p 3/2, and (c) Ga 3d core levels before and after *in situ* Ar$_{1000}^+$-ion cluster cleaning of free-standing GaN surfaces.

The O 1s core level spectra (**Figure S8**) of the bare and the monolayer AlO$_x$-coated *c*-plane GaN reveal the presence of O-H groups (even after Ar$_{1000}^+$-ion cluster sputtering), which is consistent with a hydrophilic surface (see Supporting Information S4) and the proposed reaction mechanism (**Reaction 1**).

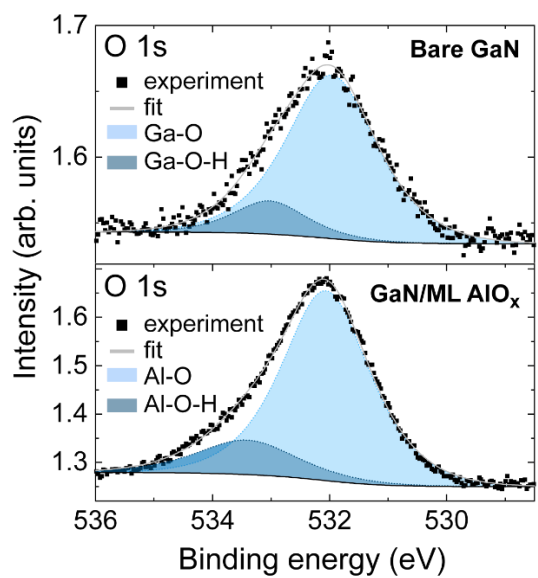

**Figure S8.** O 1s core level spectra of bare and monolayer AlO$_x$-coated GaN after *in situ* Ar$_{1000}^+$-ion cluster cleaning.

## Measurement of overlayer thicknesses by XPS

The thicknesses of the carbon overlayer, the native GaO$_x$ layer, and the monolayer AlO$_x$ film were estimated from XPS analysis (**Table S3**).

A carbon overlayer (ol) thickness, $d_{ol}$, of 7.7 ± 1.0 Å (4 samples considered) was obtained by comparing core level peak intensities before and after removal of the carbon overlayer by Ar$_{1000}^+$-ion cluster sputtering. The overlayer thickness, $d_{ol}$, can be estimated from **Equation S1**,

$$d_{ol} = -\lambda_{ol} \sin\theta \ln\left(\frac{I_1}{I_0}\right) \tag{S1}$$

where $I_0$ is the integrated core level peak intensity for a bare substrate, and $I_1$ is the corresponding core level signal attenuated by the overlayer, $\theta$ is the electron take-off angle with respect to the sample surface (here $\theta = 90°$) and $\lambda_{ol}$ is the inelastic mean free path (IMFP) of the photoelectrons in the overlayer at the kinetic energy of the measured core level (here 368 eV, Ga 2p$_{3/2}$). An IMFP for carbon ($\lambda_{ol} = 13.98$ Å) for a density of 2.2 g/cm$^3$ (graphite), computed with the Tanuma, Powell, Penn (TPP-2M) formula,[10] was obtained from the QUASES-IMFP software.

The thickness, $d_{ox}$, of the native gallium oxide layer of 10.8 ± 0.8 Å (standard deviation out of 4 different measurement spots on three samples) was obtained from XPS analysis and is in good agreement with the oxide thickness measured by XRR. The oxide thickness was estimated for the carbon-free GaN substrate (after Ar$_{1000}^+$-ion cluster sputtering) with the intensity ratio of the Ga-O and Ga-N components of the Ga 3d core level spectrum according to **Equation S2**,[11]

$$d_{ox} = \lambda_{ox} \sin\theta \ln\left(\frac{N_s}{N_{ox}} \frac{\lambda_s}{\lambda_{ox}} \frac{I_{ox}}{I_s} + 1\right) \tag{S2}$$

where $I_{ox}$ and $I_s$ are the integrated intensities of the oxide layer (ox) and substrate (s) component of the same core level (here Ga 3d). We note that in contrast to the Ga 2p peak shape, the Ga 3d peak is asymmetric such that the Ga-O and the Ga-N components can be distinguished. $N_s$ and $N_{ox}$ are the volume densities of atoms in the GaN substrate and the gallium oxide overlayer, respectively. Assuming β-Ga$_2$O$_3$ for the native oxide, the volume densities, $N = \frac{\rho}{M}$, were calculated with known material densities, $\rho$, and molar masses, $M$. Molar masses of 83.73 g/mol and 187.44 g/mol and densities of 6.1 g/cm$^3$ and 5.88

g/cm$^3$ were used to calculate the volume densities of GaN and β-Ga$_2$O$_3$, respectively. The IMFPs of the photoelectrons in the GaN substrate ($\lambda_s$ = 25.65 Å) and the Ga$_2$O$_3$ layer ($\lambda_{ox}$ = 25.27 Å) were computed with the TPP-2M formula[10] using the QUASES-IMFP software.

The AlO$_x$ thickness measured by SE (see Figure 1b) is in reasonable agreement with the thickness extracted from the relationship between the emission intensities of the Ga 2p and Ga 3d core levels. In particular, photoelectrons emitted from the Ga 2p and Ga 3d core levels probe the same element at different kinetic energies,[12] resulting in different IMFPs of $\lambda_L$ = 12.45 Å and $\lambda_H$ = 33.64 Å for photoelectrons moving through AlO$_x$ at relatively low (L) and high (H) kinetic energies, respectively. An AlO$_x$ layer thickness of $d_{ox}$ = 3.3 ± 0.8 Å was calculated for GaN exposed to 20 TMA/H$_2$ plasma cycles by using **Equation S3**,

$$d_{ox} = \frac{\ln(I_L/(I_H\,k)) \times \lambda_L \lambda_H}{\lambda_H - \lambda_L} \tag{S3}$$

where $I_L$ and $I_H$ are the integrated intensities of the Ga-N components of the Ga 2p and the Ga 3d core level spectra. To correct for the effects of carbon redeposition (Ga 3d measured after Ga 2p) and multiple layers (GaO$_x$/GaN) on photoelectron emission of the two different core levels, we introduced a compensation factor, $k = \frac{I_H^0}{I_L^0} = 1.3$, that was determined from the Ga 3d/Ga 2p intensity ratio measured for three different carbon-free and bare substrates, *i.e.*, GaN with native GaO$_x$ after Ar$_{1000}^+$-ion cluster sputtering.

**Table S3.** Overview of overlayer thicknesses obtained by XPS analysis.

| overlayer | treatment | equation | thickness, *d* (Å) |
|---|---|---|---|
| Adventitious carbon | solvent cleaned | S1 | 7.7 ± 1.0 |
| GaO$_x$ | Ar$_{1000}^+$-ion cleaned | S2 | 10.3 ± 0.9 |
| AlO$_x$ | Ar$_{1000}^+$-ion cleaned | S3 | 3.3 ± 0.8 |

We note that thickness measurements of sub-nanometer thin conformal coatings on corrugated (not atomically flat) surfaces is challenging with real-space techniques and is achieved here with SE and XPS.

Byproducts such as methane and $CH_x$ that are produced during ALD processes using TMA can potentially react with the sample surface and contaminate the ALD film.[13] However, comparison of the C 1s core level spectra before and after sputtering shows that the carbon content is just above the background noise level. The small signature that appears in the C 1s spectrum of $AlO_x$-coated GaN most likely stems from carbon redeposition, though incorporation of low concentrations of carbon into the $AlO_x$ cannot be completely ruled out (**Figure S9**). We note that substrate-enhanced growth has also been associated with precursor dissociation.[14] However, such a mechanism can be excluded in the present case since the dissociation would yield films with high carbon content. Here, XPS measurements indicate carbon signals that are indistinguishable for bare GaN and GaN coated with $AlO_x$ by thermal and $H_2$ plasma/TMA ALD (Figure S9).

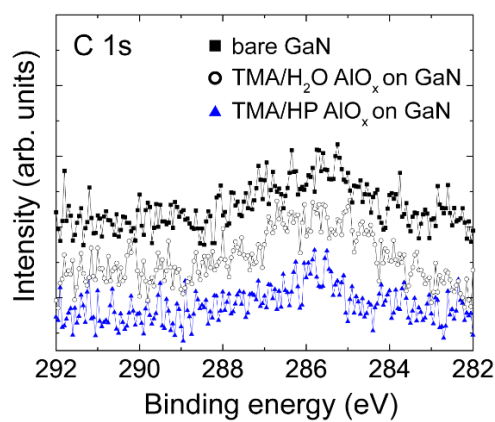

**Figure S9.** XPS of the C 1s core level after Ar-ion cluster sputtering of bare *c*-plane GaN (black squares), GaN coated with thermal $AlO_x$ following 20 cycles of TMA and $H_2O$ (black circles), and GaN coated with $AlO_x$ after 20 cycles TMA and $H_2$ plasma (blue triangles).

The XPS data shown in **Figure S10** confirm the growth of elemental aluminum during the TMA/HP process. Presumably, elemental aluminum is deposited on the GaN substrates after formation of monolayer $AlO_x$ and saturation of Ga-O-H binding sites, *i.e.* during growth regime (iii) (Figure 1). In the present work, XPS analysis of the surface without intermediate exposure to ambient conditions was not feasible and, due to the angstrom-level coverage of the hypothesized Al islands, immediate oxidation to alumina upon air exposure precluded direct verification of the presence of metallic islands. A larger $H_2$ plasma power (300 W) was used to achieve a reasonable growth rate for the deposition of a > 3 nm thick aluminum layer. Figure S10 reveals the XPS signatures of both aluminum oxide and elemental aluminum.

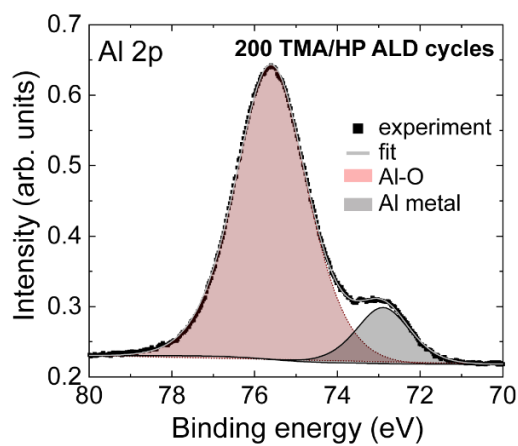

**Figure S10.** XPS of the Al 2p core level after 200 cycles of TMA and $H_2$ plasma (300 W, 2 s, 0.02 Torr) on a doped silicon substrate with a native $SiO_2$ layer.

Free-standing bare, *c*-plane GaN and bulk-like $AlO_x$ were characterized by XPS to estimate the valence band offset between monolayer $AlO_x$ and GaN using Kraut's method.[15] Therefore, a 25 nm thick (bulk-like) $AlO_x$ layer was deposited on free-standing, *c*-plane GaN by thermal ALD at 280 °C. **Figure S11** shows the analysis of valence band and core level spectra of the 25 nm thick $AlO_x$/GaN and bare GaN sample, respectively. Analysis of the spectral features attributed to surface plasmons allowed to estimate the band gap of the ALD $Al_2O_3$ film.

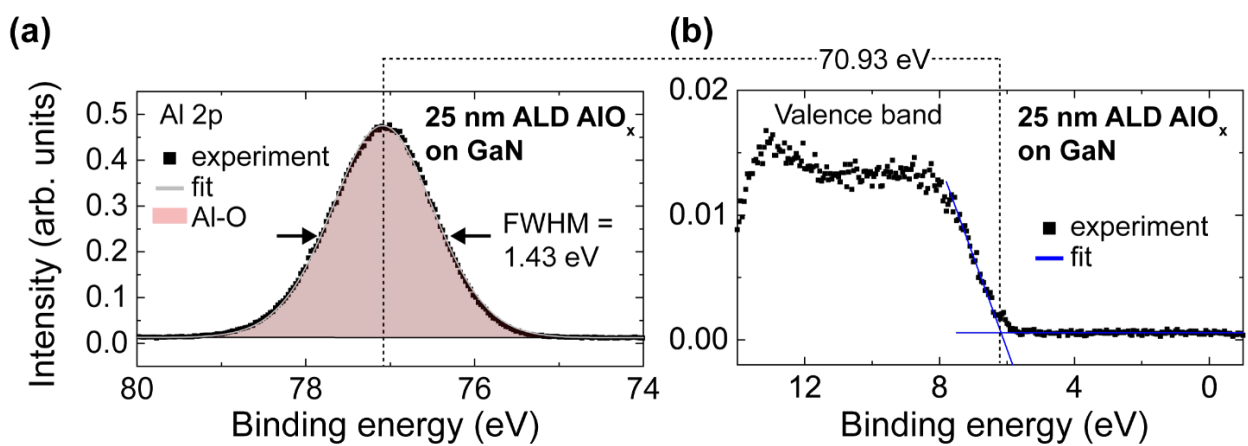

**Figure S11.** XPS spectra of the (a) Al 2p core level and (b) valence band of *c*-plane GaN coated with 25 nm thick $AlO_x$ (250 cycles of TMA and $H_2O$, 280 °C). XPS spectra of the (c) Ga 3d core level and (d) valence band of bare, *c*-plane GaN.

The valence band spectra show the characteristic signatures of GaN (**Figure S12**) where the peaks labeled as A and C at BEs of ~ 5 eV and ~ 9.5 eV are assigned as Ga 4p-N 2p and Ga 4s-N 2p bonding, respectively.[16] The spectral region labeled as "B" signifies the BE range of a GaN hybridization state. The valence band edge is at the same energy for both bare and monolayer AlO$_x$-coated GaN.

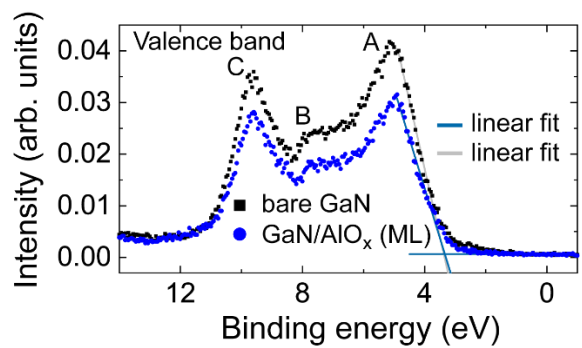

**Figure S12.** Valence band spectra of bare (black squares) and monolayer AlO$_x$-coated *c*-plane GaN (blue circles).

**Table S4** summarizes the energies of the valence band edges and the binding energies of core level components for the different sample systems.

**Table S4.** Overview of the binding energies for each of the components in the Ga 3d and Al 2p core level spectra, as well as the valence band edges, after different treatments.

| | Binding energy / eV | | | | |
| --- | --- | --- | --- | --- | --- |
| | Ga 3d $_{5/2}$ | | | Al 2p | Valence band |
| Sample | Ga-N | Ga-O | Ga-Ga | Al-O | VB edge |
| Bare *c*-plane, Ga-polar GaN | 20.45 | 21.20 | 19.01 | / | 3.37 |
| GaN/ML AlO$_x$ | 20.43 | 21.24 | 19.00 | 75.40 | 3.34 |
| GaN/ML AlO$_x$/PA-C11-OH | 20.53 | 21.26 | 19.07 | 75.60 | 3.43 |
| GaN/ML AlO$_x$/PA [1] | 20.43 | 21.15 | 18.99 | 75.43 | 3.33 |

[1] After sputtering off the alkyl chain

## S6. Atomic force microscopy of bare and monolayer AlO$_x$-coated *c*-plane, Ga-polar GaN

The power spectral density plots (**Figure S13**) show distinct peaks at 74 µm$^{-1}$ and 55 µm$^{-1}$ for the bare and ALD-treated GaN, respectively, indicating repetitive patterns that are not discerned with the frequency density analysis (Figure 2b). These patterns can be attributed to atomic steps that are aligned in a preferred crystallographic orientation and result from uniformly misoriented growth seeds, commonly observed on the surfaces of epi-ready polished GaN[17]. There are no indications of threading dislocations on the length scale characterized in this study by AFM.

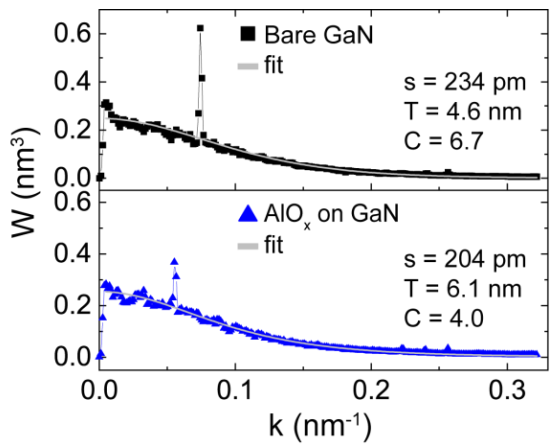

**Figure S13.** Power spectral density plots derived from the AFM images of Figure 2.

By applying a tip-sample force of 3.25 µN with diamond-like carbon (DLC) tip in contact mode, the AlO$_x$ film could be (partially) removed off the GaN substrate (Figure S14). Since DLC is harder than sapphire[18] it is suitable for abrading AlO$_x$, thereby providing another method to confirm the formation of a continuous AlO$_x$ coating and the thickness of the film (3 ± 1 Å). The estimated step height measured with tapping mode (TM) AFM is consistent with the AlO$_x$ layer thickness determined from SE measurements and in agreement with the conclusion that a single monolayer of AlO$_x$ is formed. In addition, the contrast in the TM phase image revealed local differences in the tip-sample adhesion, which is indicative of different tip-sample interaction strengths (Figure S14c) at the bare and AlO$_x$-coated GaN surfaces in the scratched and non-scratched regions, respectively. In comparison, a smaller step height and no contrast in the phase image was observed after micro scratching the bare GaN, likely due to the removal of the surface adsorbate. GaN and β-Ga$_2$O$_3$ have a hardness of 12 ± 2 GPa and ~ 9 GPa, respectively, and are expected to be harder than amorphous AlO$_x$. However, due to the atomically thin AlO$_x$ and the presence of a surface adsorbate, it is challenging to reliably determine the monolayer AlO$_x$ thickness by this micro-scratching experiment.

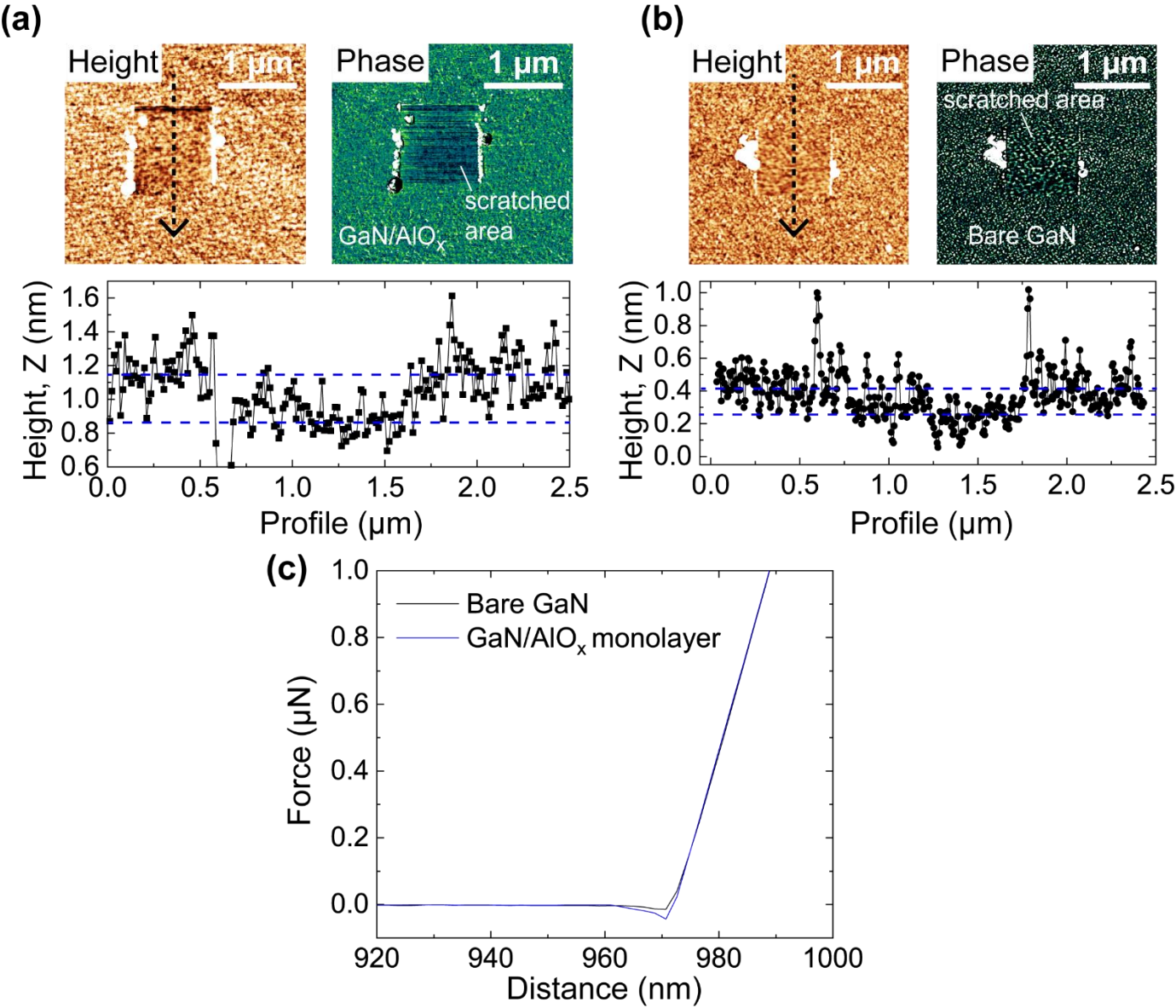

**Figure S14.** The AFM height and phase images after contact mode (3.25 µN) AFM in a 1 µm² region of a (a) monolayer AlO$_x$-coated GaN substrate and (b) a bare GaN substrate. (c) Representative force-distance curves for bare (black line) and monolayer AlO$_x$-coated GaN (blue line) substrates reveal a comparably stronger adhesion on the AlO$_x$-coated surface.

## S7. Self-assembled monolayers of phosphonic acids (SAMPs) on monolayer AlO$_x$/GaN substrates

**Preparation of SAMPs on monolayer AlO$_x$/GaN.**

SAMPs of 11-hydroxyundecylphosphonic acid (≥ 99 % (GC); referred to as PA-C11-OH) (SiKÉMIA, Montpellier, France) were prepared *via* a modified immersion technique, commonly applied for SAMPs on Al$_2$O$_3$ surfaces.[19] Due to the high surface hydrophilicity of the GaN substrates after AlO$_x$ deposition (see Table S2), wafer pieces (mounted on a Teflon sample holder) were immediately immersed vertically in a 40 ml ethanol (EtOH; VWR, Ethanol absolute ≥99.8 %, AnalaR NORMAPUR, ACS, Reag. Ph. Eur.) solution containing 2×10$^{-3}$ mol·L$^{-1}$ of the respective PA molecules. The immersion beaker was stored vibration-free in the dark under an N$_2$ atmosphere at room temperature. Although the main body of the SAMPs is formed within the first few minutes,[20] the substrates were immersed for 72 hours to ensure a complete PA surface coverage of the substrate. After immersion, the pieces were cleaned twice with EtOH in an ultrasonic bath (37 kHz, 60 W) for 5 minutes to remove weakly bound PAs, dried under N$_2$, and stored in a desiccator (~10$^{-2}$ mbar).

**XPS analysis of a SAMPs on monolayer AlO$_x$.**

A comparison between the O 1s, C 1s, P 2p, Al 2p, and Ga 3d XPS spectra of a bare monolayer AlO$_x$/GaN substrate and AlO$_x$/PA-C11-OH sample before and after Ar-ion cluster sputtering is provided in **Figure S15**. The obtained C/P ratio of 12.9 ± 0.9 deviates from the expected stoichiometry value of 11, likely due to P signal attenuation by the alkyl chain rather than from adventitious carbon. An underestimation of the P content is reasonable, since the P 2p signal intensity is attenuated, although the IMFP of P 2p photoelectrons in the organic overlayer (3.9 nm)[21] is relatively large. Correcting for attenuation by the organic overlayer is, however, difficult since the C 1s photoelectron intensity from C atoms close to the PA (head) group is more strongly attenuated compared to C atoms close to the vacuum interface. Note that the high binding energy C 1s component (~290.1 eV; Fig. S15b, black), presumably from C-O-containing adventitious carbon species adsorbed on monolayer AlO$_x$, is not perceptible on the PA-C11-OH treated substrates, indicating that the C 1s signal is only related to structural moieties of the PA itself.

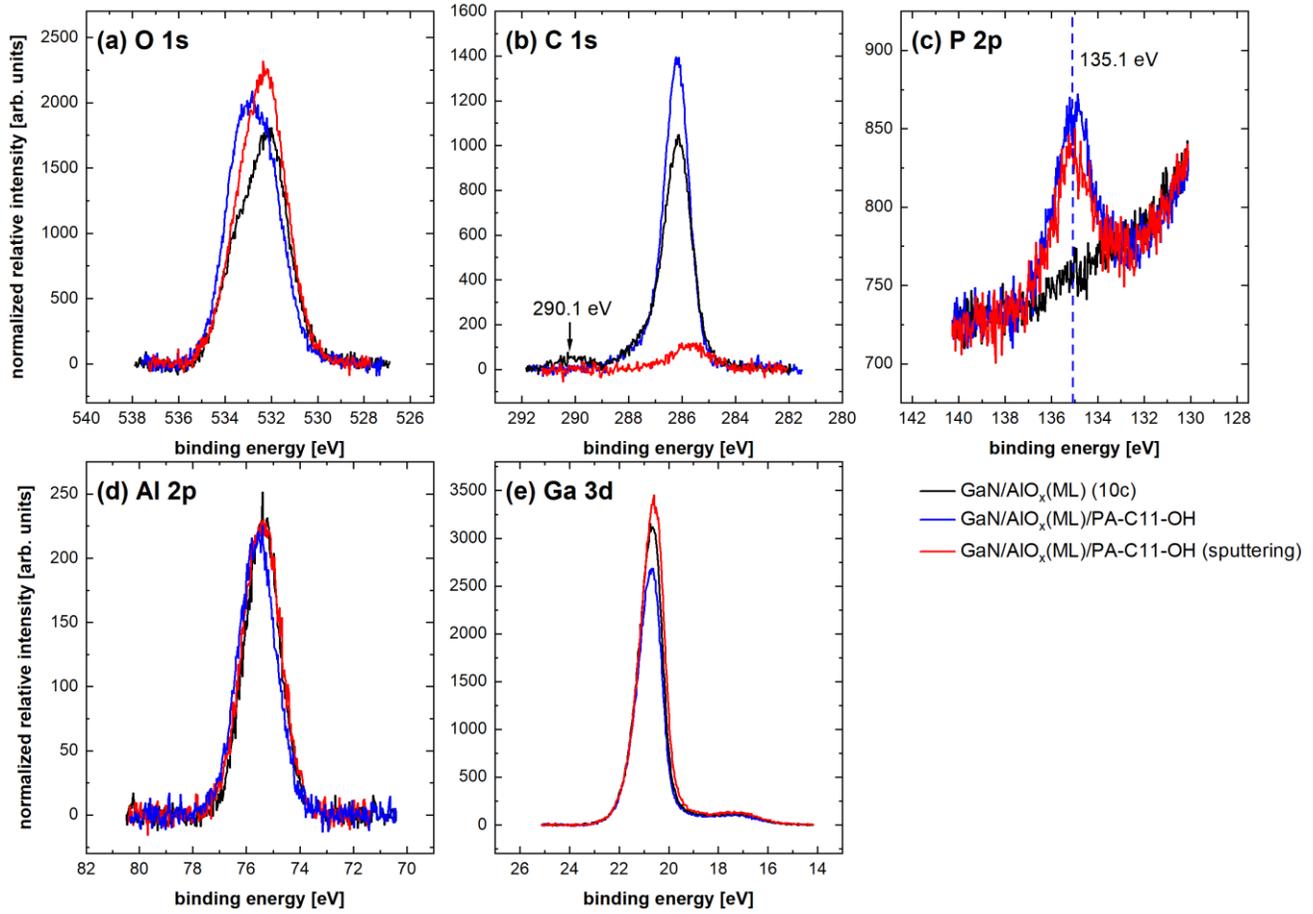

**Figure S15.** Photoelectron spectra of O 1s, C 1s, P 2p, Al 2p, and Ga 3d of a monolayer AlO$_x$ sample (black line) and an AlO$_x$/PA-C11-OH sample before (blue line) and after Ar-ion cluster sputtering (red line). Note that for the P 2p spectra no background correction was performed.

**Calculation of PA surface coverage on monolayer AlO$_x$/GaN substrates.**

The number of phosphorus atoms and, therefore, PA molecules per unit area $N_\square$ can be determined using **Equation S4**:[22]

$$N_P = \frac{A_{P2p}}{A_{Ga3d}} \cdot \frac{RSF_{Ga3d}}{RSF_{P2p}} \cdot \rho_{Ga,GaN} \cdot \lambda_{Ga3d,GaN} \cdot sin(\theta) \cdot \frac{\exp\left(\frac{d_{PA}}{\lambda_{P2p,organic}\cdot\sin(\theta)}\right)}{\exp\left(\frac{d_{PA}}{\lambda_{Ga3d,organic}\cdot\sin(\theta)}\right)} \quad (S4)$$

where $A_{P2p}/A_{Ga3d}$ is the ratio of the P 2p and Ga 3d peak areas (*i.e.*, integrated XPS intensities), $RSF_{Ga3d}/RSF_{P2p}$ (0.439/0.486) is the corresponding relative sensitivity factor (RSF) ratio (taken from the CasaXPS KratosAxis-F1s library), $\rho_{Ga,GaN}$ is the number of Ga atoms per unit volume in GaN, $\lambda_{Ga3d,GaN}$ (2.63 nm)[23] is the IMFP of Ga 3d photoelectrons in GaN (generated by Al K$_\alpha$ X-rays) $d_{\square\square}$ is the thickness of the organic overlayer, and $\lambda_{P2p,organic}$ and $\lambda_{Ga3d,organic}$ are the IMFPs of P 2p and Ga 3d photoelectrons (generated by Al K$_\alpha$ X-rays) in the organic overlayer, respectively. $\theta$ is the take-off angle of photoelectrons with respect to the sample plane ($\theta$ = 90°). Note that the RSF values presented above can only be applied for a source-to-detector geometry of 54.7°.

Neglecting surface roughness and consistent with previous studies,[24] an empirical approximation for IMFPs in self-assembled monolayers can be expressed as: $\lambda_{X,organic}$ [nm] = 0.9 + 0.0022·$E_{kin}$ [eV], where X is the photoelectron transition of interest and $E_{kin}$ [eV] is the kinetic energy in electron volts.[21] For the photoelectrons of interests, this corresponds to $\lambda_{P2p,organic}$ = 3.9 nm ≈ $\lambda_{Ga3d,organic}$ = 4.1 nm, such that the exponential term in Equation S4 can be approximated by $\dfrac{\exp\left(\dfrac{d_{PA}}{\lambda_{P2p,organic}\cdot\sin(\theta)}\right)}{\exp\left(\dfrac{d_{PA}}{\lambda_{Ga3d,organic}\cdot\sin(\theta)}\right)} \approx 1$, with an error of less than 5 % for $d_{\square\square}$ < 4 nm.[22] Since the density of the AlO$_x$ intermediate layer is unknown but likely differs from stochiometric bulk Al$_2$O$_3$, the Ga 3d photoelectrons of underlying GaN substrate were taken as a reference signal and corrected for the attenuation effect of the AlO$_x$ intermediate layer following **Equation S5**:[25]

$$A_{Ga3d} = A_{Ga3d}^{\infty} \exp\left(\frac{-d_{AlOx}}{\lambda_{Ga3d,AlOx}\cdot\sin(\theta)}\right) \tag{S5}$$

where $A_{Ga3d}^{\infty}$ is the Ga 3d peak area of a bare GaN substrate, $d_{\square\square\square\square}$ is the thickness of the AlO$_x$ intermediate layer (0.28 nm), and $\lambda_{Ga3d,AlOx}$ = 0.77 nm is an experimentally determined IMFP of Ga 3d photoelectrons in the monolayer AlO$_x$. Note that this IMFP is obtained by comparing the Ga 3d intensities of a bare (HCl-etched) to an AlO$_x$-coated GaN substrate after Ar-ion cluster sputtering. Thus, it is only strictly applicable to the system at hand. All other IMFPs were calculated based on the analysis by Werner[23a] using the TPP-2M method[23b] in NIST's database.[23c] Since the GaO$_x$ interlayer thickness is unknown, and the IMFP of Ga 3d photoelectrons in GaN (2.76 nm) and GaO$_x$ (2.75 nm) are nearly identical, $A_{Ga3d}$ in Equation S4 is integrated over all GaN- and GaO$_x$-related components. A comparison between the O 1s, C 1s, P 2p, Al 2p, and Ga 3d XPS spectra of a AlO$_x$(ML), AlO$_x$(ML)/PA-C11-OH, and AlO$_x$(ML)/PA-C11-OH (after Ar-ion cluster sputtering) is provided in Figure S15. Statistical errors are calculated by averaging the coverage obtained from 3 different pristine spots on a PA functionalized sample.

**Binding mode of PA to monolayer AlO$_x$.**

The AlO$_x$(ML) film presumably contains three different O sites.[26] O anions can act as electron-donating Lewis base sites and incompletely coordinated cations as electron-accepting Lewis acid sites, whereas OH anions can act as either a Lewis acid or base, but also as a proton-exchanging Brønsted acid-base sites. Van den Brand *et al.*[27] reported an XPS-based procedure to identify the character of O on oxide surfaces, however, for the present system higher resolution and higher signal-to-noise ratios, *e.g.* provided by synchrotron radiation, would be necessary to substantiate this claim.